\documentclass[prl,twocolumn,aps,amsfonts,nofootinbib]{revtex4}

\usepackage{epsfig}
\usepackage{graphicx}
\usepackage{amscd}
\usepackage{amsmath}
\usepackage{bm}
\input xy
\xyoption{all}

\newcommand{\beq}{\begin{equation}}
\newcommand{\eeq}{\end{equation}}
\newcommand{\bea}{\begin{eqnarray}}
\newcommand{\eea}{\end{eqnarray}}
\newcommand{\eps}{\epsilon}

\newcommand{\nn}{\nonumber}
\newcommand{\benn}{\begin{displaymath}}
\newcommand{\eenn}{\end{displaymath}}

\def\slashchar#1{\ensuremath{                               %
   \setbox0=\hbox{${}#1{}$}       
   \dimen0=\wd0                                 
   \setbox1=\hbox{/} \dimen1=\wd1               
   \ifdim\dimen0>\dimen1                        
      \rlap{\hbox to \dimen0{\hfil/\hfil}}      
      {}#1{}                                    
   \else                                        
      \rlap{\hbox to \dimen1{\hfil${}#1{}$\hfil}}   
      /                                         
   \fi}}                                        %

\begin{document}

\title{Superfluid phases of the three-species fermion gas} 
\author{Paulo F.~Bedaque} 
\affiliation{Lawrence-Berkeley Laboratory, Berkeley, CA 94720, USA} 
\affiliation{University of Maryland, College Park MD, 20742 USA}
\author{Jos\'e P. D'Incao}
\affiliation{Department of Physics, Kansas State University,
  Manhattan, KS 66506, USA } 
\preprint{}

\begin{abstract}
We discuss the zero temperature phase diagram of a dilute gas with
three fermionic species. We make use of  solvable limits  to conjecture 
the behavior of the system in the ``unitary'' regions. The physics of
the Thomas-Efimov effect plays a role in these considerations. We find
a rich phase diagram with superfluid, gapless superfluid and
inhomogeneous phases with different symmetry breaking patterns. We
then discuss one particular possible experimental implementation in a
system of $^6$Li atoms  and the possible phases arising in this
system as an external magnetic field is varied across three overlaping
Feshbach resonances. We also suggest how to experimentally distinguish
the different phases.    
\end{abstract}

\maketitle

Cold atomic gases provides are great laboratory for
many-body physics. This is due to the variety of atomic
isotopes that can be trapped and the tunability of the atomic
interactions. Another  point of interest is the ``universality'' of
some of their properties: since the detailed shape of the interatomic
potential is not probed in dilute systems, many different kinds of
interactions can be approximated by  a contact potential and different
systems present analogous characteristics. 

Our goal in  this letter is to explore  the phase diagram of a dilute
ultracold gas composed of {\it three} fermionic species~\cite{honerkamp}. It can be
considered a generalization of  
the well studied two fermionic species gas and brings about some new
features like competition among interactions and the consequences of
the Efimov efect~\cite{efimov}
Let us first summarize the main properties of the
two-fermion case as they will be useful in the more involved
three-fermion gas. There are three relevant distance scales in this
system: $r_0$ measures the range of the interatomic forces, $a$ is the
s-wave  scattering length and $1/n^{1/3}$ the typical interatomic
distance ($n$ is the particle density). We will consider only the
cases where $r_0 \ll a, 1/n^{1/3} $ since only in this case the
results have some degree of universality. The cases where $ n|a|^3 \ll
1$,  where an expansion in powers of $n|a|^3$ is possible, are more easily
analyzed.  
In the regime $ na^3 \ll 1$, $a>0$ the atoms bind into
diatomic molecules much smaller than the separation between them 
while
interacting weakly with each other. These  molecules then condense to
form a weakly coupled  superfluid. 
 If $a<0$, though, there are no molecules.
Still, the atoms form Cooper pairs and the ground
state of the system is a BCS superfluid. In the intermediate regime
$-1\alt  1/na^3 \alt 1$, the so-called ``unitary'' region, the absence
of a small parameter complicates the theoretical analysis but one can
still infer some of its qualitative properties. Notice that this
region is scaled down (in energy) version of the dilute neutron gas,
exemplifying the universality pointed out above.  In both the BEC
limit $ 1/na^3 \gg 1$ and the BCS limit $ 1/na^3 \ll -1$ the
condensation of molecules or Cooper pairs
breaks the wave function phase symmetry. That is, if we denote by
$\psi_1$ ($\psi_2$),  
the annihilation operator for the atom
of type 1 (2), the symmetry of the hamiltionian $\psi_{1,2}  \rightarrow
e^{i\alpha}\psi_{1,2}$ 
is not a symmetry of the ground state due to the
non-vanishing value of $\langle \psi_1\psi_2\rangle$. Since the
symmetry breaking pattern is 
the same on both sides of the ``unitary" region  there is not reason
to expect a qualitative change (a phase transition) in between the BEC
and the BCS limits. This picture is vindicated both by experimental
\cite{crossover_exp} and  numerical \cite{crossover_num} results.  
 
Similar arguments can be made in the three-fermion case of interest here.
There are two main new ingredients absent in the two-fermion case. The
first is the competition between the pairing/binding among different
atom pairs. 
The second is related to three-body correlations. 
Ordinarily, two-body correlations dominate the physics of dilute
systems but that may not be true in the regime $r_0/a \ll 1$. Indeed
the Thomas effect \cite{thomas} (the collapse of the three-body system
as $r_0\rightarrow 0$ at fixed $a$) and the Efimov effect
\cite{efimov} (the accumulation of three-body bound states at
threshold as $a\rightarrow\infty$ at fixed $r_0$) are consequences of
the strong attraction among three resonating particles. 
These two closely related phenomena \cite{3review} can occur only if
the three particles in the system can overlap in space, which is
forbidden in the case of only two fermion species. 
The sensitivity to the physics of three overlapping particles 
brings out a dependence of the low energy processes in a new parameter 
\cite{danilov, efimov} unrelated to two-body scattering which, in the
language of Effective Field Theories, is the value of a contact
three-body force \cite{bhk}. 
Currently, the experimental observation of the Efimov effect is a
subject of great interest due to the fundamental aspects it
represents for the three-body physics in ultracold quantum gases. 
Recently an analysis of the three-body recombination
\cite{recombination} in Bose-Einstein condensates have shown evidence
of this intriguing effect \cite{grimm_efimov}.  

 At zero temperature, with equal densities of the three
species,  the phase diagram of the system is
a function of four dimensionless variables. The dependence on the
three-body
parameter  is limited so we keep it fixed. The remaining three
parameters can be taken to be $n a_{ij}^3$, ($a_{ij}$ is the
scattering length between particles $i$ and $j$ and $n$  the density
of each atomic species). We will initially restrict our discussion to
the slightly idealized situation where two of the scattering lengths
are equal, thus assuming $a_{12}=a_{23}=a, a_{31}=\bar a$.
Our strategy to map the zero temperature phase diagram as a function
of $a$ and $\bar a$ will be to understand the system in the soluble
limits
of the phase diagram where $n|a|^3, n|\bar a|^3 \ll 1$ and
make educated guesses about how to interpolate between these
limits. This is the same successful strategy discussed above for the
two-fermion case. 

The system can be described by the hamiltonian   
\bea\label{eq:h}
\mathcal{H}&=& -\sum_{i=1}^3 \psi_i^\dagger  
\frac{\nabla^2}{2M}\psi_i+\bar g^2 |\psi_1|^2 |\psi_3|^2 \\
&+&g^2 (|\psi_1|^2 |\psi_2|^2 + |\psi_2|^2 |\psi_3|^2) + 
G |\psi_1|^2 |\psi_2|^2 |\psi_3|^2\nn .
\eea 
The couplings $g$, $\bar g$ and $G$ depend on a cutoff and are
determined respectively by $a$, $\bar a$ and, in the case of $G$, on a
three-body observable like one of the atom-dimer scattering lengths for some value of $a, \bar a$. 
The hamiltonian in eq.~(\ref{eq:h}) has a $SU_{13}(2)\times
U_2(1)\times U_{123}(1)$ symmetry, where the first factor acts only on 
atoms of type ``$1$'' and ``$3$'', the second on atoms of type ``$2$'' only and the
last is a  common phase change in all atom fields.

\begin{figure}[t]
\vskip -0.2cm
\centerline{{\epsfxsize=2.in \epsfbox{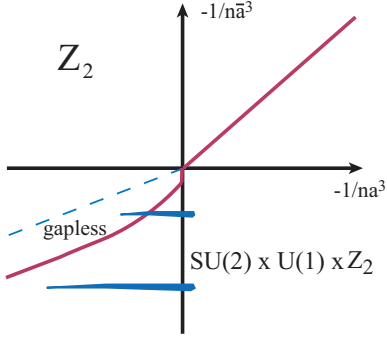}}}
\vskip -0.4cm
\caption{Conjectured phase diagram. The horizontal bands (blue online)
  show the narrow areas of atom-molecule separation. The solid (red
  online) represent the phase boundary between the $\mathbb{Z}_2$ and
  $SU(2)\times U(1)\mathbb{Z}_2$ phases. The dashed line is where a
  Fermi surface appears (Lifshitz transition).} \label{fig:phase2} 
\end{figure}  
 Let us first consider the case where both scattering lengths are
positive and $na^3, n\bar a^3 \ll 1$, located in the third
quadrant in Fig.~\ref{fig:phase2}.
 There all three pairs can form bound states of energy
$B=1/Ma^2$ or $\bar B=1/M\bar a^2$ and whose spatial extent is much
smaller than the interparticle distance. If we denote the number of
bound states between particles $i$ and $j$ by $n_{ij}$ and the number
of unbound atoms of type $i$ by $n_i$ we have that, to leading order in
$na^3$ and $n\bar a^3$, the energy of the system is given by   
\beq\label{eq:ebec}
\mathcal {E} = \frac{3(6\pi^2)^{2/3}}{10
  M}(n_1^{5/3}+n_2^{5/3}+n_3^{5/3}) - B(n_{12} + n_{23}) -\bar B n_{13}, 
\eeq which is just the sum of the molecule binding energies and the energy of the Fermi gas of unbound atoms.
We disregarded the interaction between
unbound atom-dimer, unbound atom-unbound atom 
and dimer-dimer as these contribution are, as long as their
respective scattering lengths are of the order of $\sim a,\bar a$, 
suppressed by powers of $na^3, n\bar a^3 \ll 1$ . As we
will see later the atom-dimer scattering length may diverge for some
very particular values of the parameters as a consequence of the
three-body dynamics  and  will be necessary to amend this analysis. We
find the ground state by minimizing eq.~(\ref{eq:ebec}) with the
restriction that the total number of  
atoms of each species is $n$ ($n_1 + n_{12}+n_{31} = n$ and similarly
for the other species). We find three different cases:   
\bea\label{eq:bec2}
\bar B \!&<& \!2 B :n_1=n_2=n_3=0, n_{12}=n_{23}=n_{31}=n/2, \nn \\
2B\! &<&\!\bar B < 2B + \frac{(6\pi^2n)^{2/3}}{M}: n_1=n_3=0, n_2= 
\frac{(\bar B-2B)^{3/2}}{6\pi^2}, \nn\\
\bar B\! &>&\! 2B +\frac{(6\pi^2n)^{2/3}}{M}: n_1=n_3=0, n_2=n\nn.
\eea 
This result is physically clear. 
 In the first case, it is 
favorable for all atoms to form bound states, and $n/2$ dimers are
formed between each species pair. 
All dimers species  condense leading to a non-zero value 
of the ground state expectation value $\langle \psi_i\psi_j\rangle$ for all $i,j$. 
This condensate breaks the symmetry
from $SU_{13}(2)\times U_2(1)\times U_{123}(1)$ down to
$\mathbb{Z}_2$, corresponding to a common sign flip on all atom
fields.  If $\bar B$ increases (i.e.,  $\bar a$ decreases)
beyond $\bar B > 2B$ it is favorable to  break some of the ``12" and ``23" dimers
in order to increase the number of ``31" dimers, leaving some ``2"
atoms unbound.  
The condensates 
$\langle \psi_1\psi_2\rangle$ and $\langle
\psi_2\psi_3\rangle$ will decrease in value but the symmetry breaking
pattern will remain the same. The formation of a Fermi surface for the 
``2" atoms indicates a topological (Lifshitz) phase transition
(denoted by the dashed line in fig.~(\ref{fig:phase2})).   
As we keep increasing  $\bar B$ (decreasing $\bar a$) there 
comes a point where all ``2'' atoms are unbound and only ``13'' 
dimers can be found. 
At this point $\langle \psi_1\psi_2\rangle =
\langle \psi_2\psi_3\rangle =0$ and the symmetry breaking pattern
becomes $SU_{13}(2)\times U_2(1)\times U_{123}(1)\rightarrow
SU_{13}(2)\times U_2(1)\times \mathbb{Z}_2$. The change in symmetry of
the ground state indicates the presence of a phase transition, marked
by the full line in  the third quadrant of fig.~(\ref{fig:phase2}). 

We now consider the opposite side of the phase diagram, i.e., the
region where $a,\bar a<0$, and $n|a|^3, n|\bar a|^3 \ll 1$ (first
quadrant in fig.~\ref{fig:phase2}).   In this region there are no two-body bound states, however, symmetries can be broken by the condensation of  Cooper pairs. The
most general s-wave condensate can be written as   
\beq
 \langle \psi_i \psi_j\rangle = \eps_{ijk} \mathbf{\Delta}^k,
\eeq 
where $\vec{\mathbf{\Delta}}=(\Delta_{23}, \Delta_{31}, \Delta_{12})$
is a complex vector describing the fermionic gaps. 
In the regime considered the BCS theory (mean field) can be
used. Using the  methods of reference \cite{lieb} we find the
thermodynamical potential in  this approximation to be  
\bea\label{eq:f}
 \mathcal{E}\!-\!\mu \mathcal{N}_i\!\! &=&\!\! \frac{1}{2}\int
 \frac{d^3k}{(2\pi)^3}  
 (3\eps(k) - |\eps(k)|-2 \sqrt{\eps^2(k)+|\mathbf{\Delta}|^2}\nn\\
 \!\!\!\!\!\!\!&+& \!\!
 \frac{2M}{k^2}|\mathbf{\Delta}|^2)
-M\frac{|\Delta_{12}|^2+|\Delta_{23}|^2}{4\pi
  a}-M\frac{|\Delta_{31}|^2}{4\pi\bar a}, 
\eea 
where $\eps(k)=k^2/2M -\mu$ and $\mu$ is the chemical potential. In eq.~(\ref{eq:f}) we  traded the
couplings constants by the scattering lengths. In general, in order to
maintain equal 
densities for the three species, different values 
for the  chemical potentials for the three species are required. At
low density though, those differences (of the order of $\Delta^2/k_F$,
where $\Delta$ is the gap and $k_F$ the Fermi momentum)  are
suppressed by the exponentially small size of the gap and will be
neglected here. The 
ground state is found by minimizing eq.~(\ref{eq:f}). 
The first term in eq.~(\ref{eq:f}) is a function only of the
combination $|\mathbf{\Delta}|^2$ which is a $SU_{123}(3)$ invariant. The
remaining terms don't have this symmetry and, consequently, tip the
balance in favor of one direction 
in  $\mathbf{\Delta}^k$ space. The location of the minimum then
depends on the competition between the strength of the interaction between the atoms, 
that is, on relative sizes of $a$ and $\bar a$. 
If $|\bar a| > |a|$, the energy is minimized by having  $\Delta_{13}
\neq 0$ and $\Delta_{12}=\Delta_{23}=0$. The atoms of type ``1'' and ``3'' pair up with each other while the atoms of type ``2'' remain unpaired. The symmetry breaking
pattern is $SU_{13}(2)\times U_2(1)\times U_{123}(1)\rightarrow
SU_{13}(2)\times U_2(1)\times \mathbb{Z}_2$. In the opposite case,
$|\bar a| < |a|$, the energetically favored state has   $\Delta_{13}
= 0$ and $|\Delta_{12}|^2+|\Delta_{23}|^2\neq 0$. It is equally
favorable for $\Delta_{12}$ or $\Delta_{23}$, or a combination of
them to be non-zero, due to the $SU_{13}(2)$ symmetry of the
hamiltonian. The symmetry breaking pattern is $SU_{13}(2)\times
U_2(1)\times U_{123}(1)\rightarrow  \mathbb{Z}_2$.  In this phase
atoms of type ``1'' and ``3''  
do not pair up with each other but 
pair up
with atoms of type ``2''. As
a consequence of the different pairing patterns we conclude that there
must be a phase transition along the line $a=\bar a$ in the first
quadrant.  Both phases are superfluid since both contain Cooper pairs
but the $SU_{13}(2)\times U_2(1)\times \mathbb{Z}_2$ phase also has
metal-like properties on account of the unpaired atoms of type ``2''. 
%

Let us consider now the second quadrant in (Fig.~\ref{fig:phase2}),
approaching it from the third quadrant by moving along the vertical  
direction. As we saw, in the third quadrant just below the horizontal
axis all atoms are bound. 
As we cross  the horizontal axis, atoms of type ``1'' and ``3'' are no
longer bound but can still form Cooper pairs. In fact, the transition
between bound atoms and Cooper pairs does not change the symmetries of
the ground state (that remains $\mathbb{Z}_2$) and no phase transition
is expected. This is perfectly analogous to the BEC/BCS transition
discussed of the two-fermion system.  Finally, let us consider the
fourth quadrant. There is also a smooth connection between third and
fourth quadrants as $ a$ flips sign. In the third quadrant, just to
the left of the vertical axis, atoms ``2'' are attracted to the other
ones, but all atoms ``1'' and ``3'' are already bound and not
available for binding with ``2'' and the attraction to atoms 
``2''
is not strong enough to force a rearrangement of the binding. As we
move horizontally towards the right and cross the vertical  axis, the
interaction of atoms of type 
``2''
with atoms ``1'' and ``3'' becomes weaker and does not even
support a bound state. Consequently, no qualitative change occurs and
no phase transition is expected. 
The symmetry breaking pattern remains the same and no phase transition
is expected. 

Up to now we have taken into account only two-body correlations -- the
existence of two-body bound states and/or Cooper pairs -- to discuss
the ground state of the system. This is reasonable in dilute systems
with short range interactions as the probability of having more
particles at the same point in space is small. Our previous discussion
receives corrections  proportional to $n|a|^3, n|\bar a|^3\ll 1$  and
is, consequently, robust. This argument however, implicitly assumes
that parameters describing the three-body physics, for instance, the
scattering length between an atom and a molecule, are of the same
order as $a$ or $\bar a$. The presence of a three-body scale
,however,
may invalidate the argument. This is indeed what happens for some
specific values of the parameters. It is known that the atom-dimer
scattering length becomes arbitrarily large for specific values of the
parameters. When the atom-dimer interaction is 
 strong
and repulsive the possibility of spatial separation between atoms and
molecules arises. Free atoms and molecules appear simultaneously only
in the third quadrant of the phase diagram and only the atom of type
``2'' is found unbound among  molecules of type ``13''.   
The criterion for the (linear) stability against atom/molecule
separation in a  boson-fermion (atom-molecule) mixture is \cite{pethick}  
\beq\label{eq:stability}
\frac{ (6\pi^2)^{5/3}}{9\pi }\frac{a_{13,13}}{n_2^{1/3}} > 9\pi^2
a_{13,2}^2, 
\eeq 
where $a_{13,13}\approx 0.6 \bar a$ \cite{petrov} is the scattering
length between two ``13'' dimers and, for 
$\bar a < a$ 
we have
\cite{efimov} 
\beq\label{eq:a3}
 a_{13,2} = \bar a (A+B \tan (s_0 \log\frac{\bar a}{r_0}+\Phi)),
\eeq 
where $A, B$ and $ s_0\approx 1.006$ are numerical constants. Unfortunately,
$\Phi$ is a function of the three-body force $G$ and thus can only be
determined with the knowledge of a three-body observable. 
 For $a<\bar a$,
there is no enhancement and $ a_{13,2}\sim \bar a$. 
 Notice that a such divergence of the atom-dimer scattering length (the 
poles in Eq.~(\ref{eq:a3})) only occur for fermionic gases if there
is, at least, three different species. 
In other cases the Pauli principle forbids the atoms to overlap and
limits the strength of the interaction among the three particles. 
For $\bar a$ near one of the poles of Eq.~(\ref{eq:a3}) the stability
condition given by Eq.~(\ref{eq:stability}) is violated and atoms of
type ``2'' separate from the ``13'' molecules.
%
%
%
%
The experimental observation of this phenomenon, as well as its
repetition for different poles in $a_{13,2}$, would provide a striking
 confirmation of the Efimov effect.   
%
%
The bands where this instability occurs are repeated every time $\bar
a$ is increased by a factor $e^{\pi/s_0}\approx 22.7$. We have
indicated two of these bands in our phase diagram in
Fig.~\ref{fig:phase2}. Their fate as we cross 
the vertical axis is  unclear, i.e., they may or may not extend into
the fourth quadrant. The precise position where these bands occur when
we approach the unitary regime ($-1/n|a|^3\ll 1$) varies for each 
systems as it depends on three-body information encapsulated in $G$. 
It is hard to determine whether this repetition continues to very
small values of $-1/n|a|^3$ and we then enter the unitary regime where
our methods fail. 

%
Our considerations can be easily extended for the case where all three
atom-atom scattering lengths are different. 
Here we will consider the phases arising in a concrete possible
realization of the three-fermion gas. 
 It was predicted in Ref.~\cite{grimm} the existence of three Feshbach
resonances between atoms in the lowest three hyperfine states of
$^6$Li for values of the magnetic field ${\bf B}$ in the range of 
$60$ mT
$<{\bf B}< 100$ mT.   
In this case, all scattering lengths are positive for small 
fields and negative for large fields. 
The resonances occur around ${\bf B}\approx 69$ mT (``13''), ${\bf B}\approx 81.1
$mT and  (``12'') and ${\bf B}\approx 83.4$ mT (``23''), where the respective
scattering length diverges and flip sign. For the largest ${\bf B}$ in
this range all $a_{ij}$ are negative, the ground state of the system 
 is an analogue of the phase in the first quadrant 
in fig.~(\ref{fig:phase2}) and the condensate is driven by the
largest scattering length in absolute value. A
$\langle \psi_1\psi_2\rangle$ condensate forms which breaks the
$U_1(1)\times U_2(1)\times U_3(1)$ symmetry of the hamiltonian down to
$U_{1-2}(1) \times U_3(1)\times \mathbb{Z}_2$ ($U_{1-2}(1)$
corresponds to an equal but 
opposite phase rotation of the atoms ``1" and ``2"). As the field is
lowered below ${\bf B}=81.1$ mT, $a_{12}$ and $a_{23}$ flip sign but
that initially does not change the phase of the 
system.  Atoms ``1" and ``2" that formed a Cooper pair 
 at large ${\bf B}$ 
now form a bound state. Lowering $B$ further still, at some point  the
binding energies 
for the pairs ``12'' and ``23''
are close to each other and
large enough to dominate the energetics of the system. At this point,
type ``12'' and ``23'' molecules form and condense. The creation of
condensates $\langle \psi_1\psi_2\rangle$ and $\langle
\psi_2\psi_3\rangle$ break the symmetry down to $U_1(1)\times
U_2(1)\times U_3(1)\rightarrow \mathbb{Z}_2$, indicating the existence
of a different phase from the one at high values of ${\bf B}$. This is
the analogue of the phase transition in the third
quadrant of fig.~(\ref{fig:phase2}).  
As we mentioned above,
this picture may be modified by the atom-molecule separation for
certain values of ${\bf B}$ below the ``12" resonance, but we cannot
predict for which value of  ${\bf B}$ this will occur due to our
ignorance of the parameter $\Phi$ (or $G$). 
%
The phase separation would be observed in practice if one would be
able to scan $a_{12}$ 
through a factor of $e^{\pi/s_0}\approx 22.7$ in order to be sure to
cross one of instability bands.
%
%
A measurement of three-body observables involving these three
states of $^6$Li would determine $G$ (and $\Phi$) and would allow us
to make a more quantitative prediction.  

We close by pointing out that the two different superfluid phases can
be distinguished by the properties of their vortices. In the three
$^6$Li system, for instance, and at high fields, atoms of type ``3''
are unpaired and should not participate in in the formation of
superfluid vortices. For lower fields, below the point where the phase
transition occur all atom types can only rotate by the formation of
vortices. The creation of these vortices by  means of probes coupling
to only one atom type is a direct way of exploring the phase diagram
we have discussed.

%
%
  
\end{document}